\documentclass[conference]{IEEEtran}
\IEEEoverridecommandlockouts
\usepackage{cite}
\usepackage{amsmath,amssymb,amsfonts}
\usepackage{algorithmic}
\usepackage{graphicx}
\usepackage{textcomp}
\usepackage{xcolor}
\def\BibTeX{{\rm B\kern-.05em{\sc i\kern-.025em b}\kern-.08em
    T\kern-.1667em\lower.7ex\hbox{E}\kern-.125emX}}

\usepackage{amsmath,amsfonts,amsthm,physics}    
\usepackage{graphicx}                           
\usepackage{placeins}                           
\usepackage{xcolor}                             
\usepackage{comment}                            
\usepackage{multirow}                           
\usepackage{blindtext}                          
\usepackage{siunitx}                            
\usepackage{float}                              
\usepackage{lipsum}                             
\usepackage{tikz}                               
\usetikzlibrary{positioning}                    
\usetikzlibrary {shapes.geometric}              
\usepackage{mathtools}                          
\usepackage{svg}                                
\usepackage{soul}                               
\usepackage[symbol]{footmisc}

\newcommand{\e}{\varepsilon}

\begin{document}
\onecolumn
\null
\vfill
\begin{center}
    \huge{This article has been accepted for publication in the IEEE International Conference on Quantum Communications, Networking, and Computing 2024. This is the accepted manuscript made available via arXiv. }
\end{center}
\vfill
\normalsize{© 2024 IEEE. Personal use of this material is permitted. Permission from IEEE must be obtained for all other uses, in any current or future media, including reprinting/republishing this material for advertising or promotional purposes, creating new collective works, for resale or redistribution to servers or lists, or reuse of any copyrighted component of this work in other works.}

\twocolumn
\clearpage

\bstctlcite{IEEEexample:BSTcontrol}
\title{Quantum-Amplified Simultaneous Quantum-Classical Communications
}
\author{\IEEEauthorblockN{Nicholas Zaunders\textsuperscript{1,2}, Ziqing Wang\textsuperscript{1}, Timothy C. Ralph\textsuperscript{2}, Ryan Aguinaldo\textsuperscript{3}, and Robert Malaney\textsuperscript{1}}\\
\IEEEauthorblockA{
\textit{$^1$School of Electrical Engineering and Telecommunications, University of New South Wales, Sydney, NSW 2052, Australia.}\\
\textit{$^2$Centre for Quantum Computation and Communication Technology, School of Mathematics and Physics,}\\ 
\textit{University of Queensland, St Lucia, Queensland 4072, Australia.}\\
\textit{$^3$Northrop Grumman Corporation, San Diego, CA 92128, USA.}
\vspace{-10pt}
}}
\maketitle

\begin{abstract}
Classical free-space optical (FSO) communication promises massive data throughput rates relative to traditional wireless technologies - an attractive outcome now being pursued in the context of satellite-ground, inter-satellite and deep-space communications. The question we investigate here is: how can we minimally alter classical FSO systems, both in infrastructure and in energy input, to provide some element of quantum communication coexisting with classical communications? To address this question, we explore additional Gaussian displacements to classical FSO encoding on the satellite, determining the minimum signal requirements that will meet given specifications on the combined classical and quantum communications throughput. 
We then investigate whether enhanced quantum-based amplifiers embedded in receivers, which have proven advantageous in standalone quantum communication, can enhance our combined classical-quantum communication throughput. We show how this is indeed the case, but only at the cost of some additional receiver complexity, relative to standalone quantum communications. This additional complexity takes the form of an additional beamsplitter and two heterodyne detectors at the receiver. Our results illustrate a viable pathway to realising quantum communication from classical FSO systems with minimal design changes.
\end{abstract}

\section{Introduction} \label{sec:intro}
Quantum key distribution (QKD) protocols promise unconditional information-theoretic security through the fundamental constraints of quantum mechanics \cite{bennett_quantum_1992, shor_simple_2000}. Of these protocols, those exploiting continuous variables (CV-QKD) to communicate secret-key information show the most promise for economical and near-term deployment. These protocols, the most ubiquitous of which is the Gaussian-based GG02 scheme introduced in \cite{grosshans_continuous_2002}, encode information into the phase quadratures of semiclassical optical coherent states.

An advantage of CV-QKD protocols is their ability to simultaneously carry both the weak modulation required to encode secret-key information as well as a larger separate modulation, which can be exploited to encode a classical digital signal \cite{qi_noise_2018}. This benefit is unique to CV-QKD, owing to the semiclassical nature of the quantum states being communicated, and offers the utility of being able to append secret-key information onto classical signals, or vice-versa, without incurring a major penalty to either signal's performance. This method of simultaneous encoding is referred to as simultaneous quantum-classical communication (SQCC). This scheme has been experimentally realised \cite{kumar_experimental_2019} as well as extended to more sophisticated MDI CV-QKD protocols \cite{pan_simultaneous_2020, hu_simultaneous_2022}.

However, while CV-QKD offers ease of use, contemporary protocols are heavily restricted by poor loss tolerance, with state-of-the-art experimental implementations unable to deliver a secret key above 32.5 dB of channel loss \cite{zhang_long-distance_2020}. Recent works analysing the theoretical behaviour of SQCC protocols, which possess extra noise components as a result of the appended classical signal, have described one-way losses of up to 16~dB under realistic conditions \cite{qi_noise_2018}. In the context of satellite-optical communication, the low loss tolerance of SQCC CV-QKD protocols represents a major barrier to effective deployment. Additionally, all previous studies of combined classical-quantum systems \cite{qi_simultaneous_2016, qi_noise_2018, kumar_experimental_2019, pan_simultaneous_2020, hu_simultaneous_2022} obtain results relying on an implicit freedom in the available energy of their systems, a feature that is not guaranteed in the regimes in which satellite platforms operate as a result of practical energy-usage constraints.

One potential loss mitigation strategy is receiver-side noiseless linear amplification. Noiseless linear amplifiers (NLAs) amplify arbitrary quantum states by some gain $g \geq 1$ with a finite probability of success; in the limiting case, an NLA performs perfect amplification $\ket{\alpha} \longrightarrow \ket{g\alpha}$ with a zero probability of success \cite{ralph_nondeterministic_2009}. Several physically-realizable noiseless amplifier designs, which approximate the above ideal case for low-amplitude states, have been proposed, including hardware-based implementations \cite{xiang_heralded_2010, blandino_improving_2012, jing_improving_2021, ghalaii_long-distance_2020, notarnicola_long-distance_2023} as well as virtual measurement-based NLAs \cite{chrzanowski_measurement-based_2014, fiurasek_gaussian_2012}. Recent studies \cite{ghalaii_long-distance_2020, notarnicola_long-distance_2023} of CV-QKD protocols employing receiver-side NLA demonstrate a substantial increase in protocol range, with keyrates scaling proportionally to the optimal repeaterless bound \cite{notarnicola_long-distance_2023}. 

In this work, we explore the feasibility of satellite-based SQCC. Our novel contributions are as follows. 
\begin{enumerate}
  \item We provide analytical results predicting the minimum beam intensity needed to achieve arbitrary classical and quantum communication quality-of-service thresholds, for a given channel loss and protocol characteristics.
  \item We examine the feasibility of performing SQCC via satellite platforms by evaluating and optimizing communications performance in the low-photon-number regime, for a novel protocol utilising quantum amplification at the receiver side. We demonstrate that the use of physically-realizable amplification techniques provides a strong advantage over non-amplified protocols.
  \item We propose a scheme which combines receiver-side quantum amplification with a novel dual-measurement strategy and optimize this protocol over free parameters. We show that in the ideal case such a protocol provides near-optimal quantum and classical performance.
\end{enumerate}
Our results show that, when utilised in tandem with tools such as quantum amplifiers, SQCC protocols offer an energy-efficient and compact framework through which practical integrated quantum-classical networking can be achieved.

\section{Protocol specifications and energy features of SQCC CV-QKD} \label{sec:sqcc}

In this section we investigate the behaviour of the SQCC protocol proposed in \cite{qi_simultaneous_2016} in relation to quantum and classical quality-of-service metrics. We determine the beam photon-number requirements associated with achieving a given minimum quality-of-service threshold and subsequently identify two distinct operating regimes for which the protocol obtains feasible secret-key and classical throughput. 

The protocol in the entanglement-based picture is as follows. Alice first generates a zero-mean two-mode squeezed vacuum state (TMSVS) of squeezing parameter $0 \geq \lambda \geq 1$. The TMSVS is Gaussian in both modes \cite{weedbrook_gaussian_2012} and is therefore fully characterised by the state's first (mean) and second (covariance) statistical moments in the optical phase space. The state is thus completely represented by the covariance matrix
\begin{align} \label{tmsvs-cv-prechannel}
    V_{AB} &= \begin{pmatrix}V\mathbb{I}&\sqrt{V^2 - 1}\sigma_z\\\sqrt{V^2 - 1}\sigma_z&V\mathbb{I}\end{pmatrix},
\end{align}
where $\mathbb{I}$ is the 2x2 identity matrix, $\sigma_z$ the Pauli matrix in the $Z$-basis, and $V = \frac{1 + \lambda^2}{1 - \lambda^2}$ the variance of Alice and Bob's modes~\cite{weedbrook_gaussian_2012}. All variances are expressed in shot-noise units. 

Upon generation of the state, Alice performs heterodyne detection on one mode, labelled $A$. Alice then encodes a series of classical bits on the remaining TMSVS mode, labelled $B$, by effectively shifting the mean value of her Gaussian modulation. In the prepare-and-measure scheme, this corresponds to sending an ensemble of states $\ket{\Tilde{\alpha} + q}$, where $q \sim \mathcal{N}(0, V)$ is a random complex value drawn from a zero-mean bivariate Gaussian distribution of variance $V$. The phase-space displacement $\Tilde{\alpha}$, corresponding to the mean of the overall distribution of states, is drawn from the alphabet of a classical coherent-state optical communications protocol, e.g. BPSK, where $\Tilde{\alpha}$ is drawn from $\{ +\alpha_C, -\alpha_C : \alpha_C \equiv \alpha e^{i\theta} \}$. In the entanglement-based scheme, Alice performs this encoding by displacing her transmitted mode $B$ appropriately.

Alice then sends the mode $B$ to Bob via a thermal-loss channel with transmissivity $T$ and equivalent thermal excess-noise contribution $\e$. This channel is a Gaussian map~\cite{laudenbach_continuousvariable_2018}, which in general sends a pure Gaussian TMSVS with covariance matrix \eqref{tmsvs-cv-prechannel} to a Gaussian thermal state described by covariance matrix~\cite{weedbrook_gaussian_2012}:
\begin{align}\label{tmsvs-cv}
    V_{AB} &= \begin{pmatrix}V\mathbb{I}&\sqrt{T(V^2 - 1)}\sigma_z\\\sqrt{T(V^2 - 1)}\sigma_z&T(V+\chi)\mathbb{I}\end{pmatrix}.
\end{align}
The quantity $\chi = \frac{1 - T}{T} + \e$ represents the total channel noise resultant from the transmission loss and thermal bath \cite{lodewyck_controlling_2005}.

At the receiver end, the signal is measured via heterodyne detection. To extract the classical information, Bob performs a threshold discrimination on the measured values depending on the agreed-upon classical scheme. Given the classical bits, Bob can then extract the random Gaussian information by subtracting the displacement associated with the measured classical bit, reversing Alice's displacement and reverting the protocol to an equivalent zero-mean Gaussian-modulated coherent-state (GMCS) CV-QKD protocol \cite{qi_simultaneous_2016}. (Here we would like to note that schemes exploiting time-multiplexing of separate classical and quantum signals present an alternative pathway to combined quantum-classical communications, compared to the simultaneous encoding described above. However, these schemes are not covered in this work.)

For simplicity, our protocol adopts a BPSK classical communications scheme, where Alice's classical information is encoded as a large complex displacement in the $+\alpha$ or $-\alpha$ direction in the phase space, corresponding to a 0 or 1 respectively. In the entanglement-based scheme, this requires Alice to displace the outgoing mode of her TMSV state by $\pm \alpha$ depending on which classical bit she wishes to encode.

We now proceed to calculate the effective secret key rate and classical communications efficiency of the SQCC protocol. Firstly, we choose to model the way in which the additional classical communication affects the quantum scheme by introducing additional sources of untrusted Gaussian excess noise~\cite{qi_simultaneous_2016}, where we write $\e \equiv \e_0 + \e(\alpha)$ for amplitude-independent excess noise $\e_0$ and amplitude-dependent noise $\e(\alpha)$ arising from the classical encoding. We choose to adopt a simple model where the amplitude-independent noise is identified with the thermal noise deriving from the channel, and the amplitude-dependent noise is given by $\e(\alpha) \equiv \e_\sigma = \alpha^2 \sigma$ \cite{qi_noise_2018}. This term describes the phase-instability noise associated with large coherent states, for coherent displacement of magnitude $\alpha$ and phase-noise coefficient $\sigma$, given $\alpha^2 \geq V$. 

We can compute the bit-error rate (BER) $e_C$ of Bob's classical information via\cite{qi_simultaneous_2016}
\begin{align}
    e_C &= \frac{1}{2} \text{erfc} \left( \sqrt{\frac{T\alpha^2}{2 B}} \right),
\end{align}
where the total noise variance $B$ of Bob's received signal prior to measurement and postprocessing is given by
\begin{align}
    B &= T(V + \chi) \notag\\
    &= T(V - 1 + \e_0 + \e_\sigma) + 1.
\end{align}
Given the BER, we also observe that Bob's quantum signal is susceptible to a second type of amplitude-dependent untrusted noise, specifically the noise $\e_{BER} = 4\alpha^2 e_C$  \cite{qi_noise_2018} introduced by performing the wrong postprocessing displacement when the classical bit is measured incorrectly.
We thus compute the total covariance matrix of Bob's quantum measurement results including effects from both channel noise and postprocessing associated with the SQCC protocol:
\begin{align}\label{tmsvs-cv-sqcc}
    V_{AB}^{SQCC} &= \begin{pmatrix}V\mathbb{I}&\sqrt{T(V^2 - 1)}\sigma_z\\\sqrt{T(V^2 - 1)}\sigma_z&T[V+\chi + \e_{BER}]\mathbb{I}\end{pmatrix}.
\end{align}

\begin{figure}[!b]
    \centering
    \vspace{-1.5em}
    \includegraphics[width = 0.85\columnwidth]{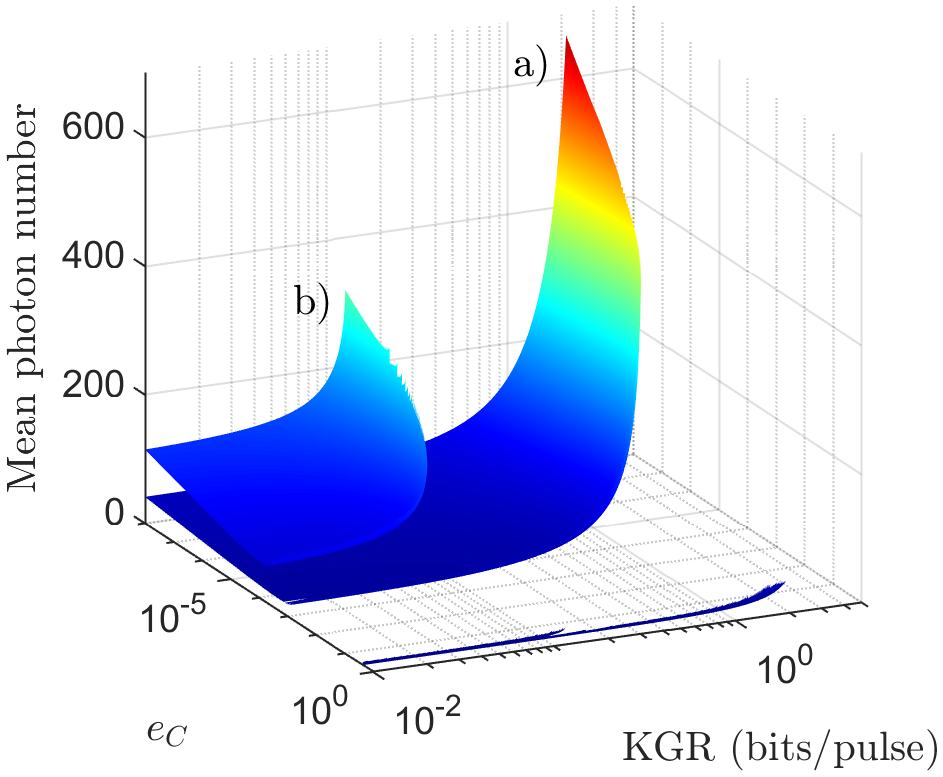}
    \vspace{-3mm}
    \caption{Minimum pulse photon number required to achieve a given protocol bit-error rate and secure key generation rate (KGR) for BPSK Gaussian-modulated coherent-state SQCC CV-QKD operating over a channel of a)~5~dB and b) 0 dB loss. We define channel excess noise $\e_0 = 0.03$, quantum reconciliation efficiency $\beta = 0.95$ and phase noise $\sigma = 10^{-6}$. For each channel, two distinct regions are seen: a large-$\alpha$ regime characterised by large mean photon numbers and bit-error rates below $10^{-4}$, and a small-$\alpha$ regime (thin blue region) characterised by ultra-low photon numbers, comparable secret-key rates to the large-$\alpha$ regime and bit-error rates approaching $10^0$.}
    \label{fig:manyalpha-combined-sqcc-vopt}
\end{figure}

Bob's final measurement results thus take the form of a set of zero-mean Gaussian-distributed random variables $p_{B}(q_B,p_B)$ correlated with Alice's heterodyne measurement outcomes $p_{A}(q_A,p_A)$, as in ordinary GMCS CV-QKD. Alice and Bob then perform reverse reconciliation and privacy amplification on their joint distribution, thereby extracting a secret key \cite{bennett_generalized_1995}. In the reverse reconciliation scheme, which is optimal for QKD \cite{pirandola_advances_2020}, the rate $K$ at which a secret key is extracted in the asymptotic regime is written as \cite{laudenbach_continuousvariable_2018}
\begin{align}
    K &\geq \beta I_{AB} - \chi_{EB},
\end{align}
for reconciliation efficiency $0 \geq \beta \geq 1$, Shannon mutual information derived from the shared joint variables $I_{AB}$, and Holevo information $\chi_{EB}$ representing the maximal information Eve can extract from her attack on the protocol. Furthermore, we make the assumption that the eavesdropper Eve has total control of the channel characteristics $\{ T, \e_0 \}$, and is able to perform collective attacks, which represent the best possible strategy in the asymptotic limit \cite{grosshans_collectiveattacks_2005}. 

Given this, the covariance matrix of Alice and Bob's joint quantum state given by Eq. \eqref{tmsvs-cv-sqcc} provides sufficient information to fully calculate the key generation rate $K(V, T, \e, \alpha)$ as a function of Alice's initial TMSV state variance, chosen classical displacement, and the channel parameters. Thus, for a generic thermally-mixed two-mode squeezed state described by the general covariance matrix
\begin{align}
    V_{AB} &= \begin{pmatrix} A\mathbb{I} & C\sigma_z \\ C\sigma_z & B\mathbb{I} \end{pmatrix},
\end{align}
we write 
\begin{align}
    I_{AB} &= \log_2\left( \frac{A + 1}{A + 1 - \frac{C^2}{B + 1}} \right),\\
    \chi_{EB} &= G(\nu_1) + G(\nu_2) - G(\nu_3).
\end{align}
Here $\nu_{1,2}$ and $\nu_3$ are the symplectic eigenvalues of the covariance matrices $V_{AB}$ and $V_{A|B}$, corresponding to Alice and Bob's joint state and Alice's state conditional on Bob's measurement respectively. Lastly, we choose to evaluate the key rate $K(V, T, \e_0, \alpha)$ as a function of transmission loss $\mathcal{L}$ in dB to facilitate comparison with free-space channels. 

We now proceed to investigate the energy landscape of the SQCC protocol established above.
We consider the mean photon number $\overline n$ of the pulse, which is dependent on both the classical displacement $\tilde \alpha$ as well as Alice's modulation variance $V$. For the BPSK-based SQCC protocol described above, the mean photon number is given by $\overline n(\alpha, V) = \alpha^2 + 2V$. If the channel characteristics $\{T, \e_0\}$ are fixed, then the two free parameters $\alpha$ and $V$ also generate the two protocol quality-of-service metrics, which we define as the secure quantum key bits exchanged per pulse $K(\alpha, V)$ and classical bit-error rate $e_C(\alpha, V)$. 

Interestingly, we observe that for any given pair of quantum and classical performance targets $\{K_0, {e_C}_0\}$, the pulse strength required to achieve said targets is not necessarily uniquely defined: for some regions of the parameter space, Alice may employ multiple strategies, corresponding to multiple pairs of $\{\alpha, V\}$, to achieve the same bit-error rate and secure key rate for different energy costs. Figure \ref{fig:manyalpha-combined-sqcc-vopt} therefore highlights the relation between these quality-of-service metrics and the \textit{minimum} mean photon number required to achieve them. We find that the overall cost is dominated by the classical displacement $\alpha^2$, with the minimum photon number increasing as the requirement on classical communications quality is increased. A much sharper increase in the minimum photon count is also seen as quantum performance requirements increase, which we attribute to the nonlinear (singly-peaked) amplitude-dependent excess noise contribution $\epsilon_{BER}$, which requires either a very large or very small $\alpha$ to minimise.

As a consequence, for each channel we also observe two distinct and separate regions of the parameter space for which the protocol can feasibly operate. One region, which we identify as the large-$\alpha$ regime, forms the bulk of the results and is characterised by low ($e_C < 10^{-4}$) classical bit-error rates and relatively high energy requirements ($\overline n\sim10^2$). This region is where the majority of work in this field has operated \cite{qi_simultaneous_2016, qi_noise_2018, pan_simultaneous_2020}. However, we identify a new region, present as a thin line along $e_C \sim 10^0$, which we identify with a hitherto-unseen small-$\alpha$ regime. This region is characterised by competitive quantum performance on par with the large-$\alpha$ regime and low classical performance, and can also operate in the ultra-low-photon-number regime ($\overline n\sim10^0$).

\section{SQCC CV-QKD with receiver-side NLA} \label{sec:sqcc-nla}
We now proceed to evaluate the performance of the SQCC CV-QKD protocol described above when noiseless amplifiers are used at the receiver side. We investigate the effect of the ideal NLA and quantum scissor-type NLA on the secret key rate as a function of channel loss $\mathcal{L}$ and classical amplitude $\alpha$ and optimize performance over modulation variance $V$ and gain $g$. We also summarise the classical performance of each protocol. We then propose a novel scheme exploiting separated classical and quantum measurement stages and evaluate communications performance relative to previous results.

\subsection{Ideal NLA} \label{sec:sqcc-idnla}
The simplest model of noiseless amplification within a CV-QKD protocol is the ideal NLA. The ideal NLA operation is described by the operator $\hat g = g^{\hat n}$, where $\hat n$ is the Fock-basis number operator $\hat n = \hat a^\dagger \hat a$ and $g \in \mathbb{R}^+$ is some arbitrary real number describing the gain of the amplifier, such that the operator performs the ideal amplification $\hat g \ket{\alpha} \rightarrow \ket{g\alpha}$. This operation is unbounded and cannot occur deterministically without violating the no-cloning bound \cite{ralph_nondeterministic_2009}; however, it may be performed non-deterministically, i.e. with some finite chance of amplification. The probability of successful amplification vanishes exponentially fast as any operator $\hat g$ approaches $g^{\hat n}$; we therefore make the general assumption that the probability of success $P^{ID}$ in the ideal case is equivalent to $1/g^2$, since this represents an upper bound \cite{blandino_improving_2012}.

For Gaussian coherent-state-based protocols such as the SQCC protocol described in Section \ref{sec:sqcc}, a protocol utilising ideal amplification at the receiver side is equivalent to a non-amplified protocol with modified effective channel parameters. The key rate of the amplified protocol is thus written as
\begin{align}\label{KGR-SQCC-ID}
    K^{ID}(g, V) = \frac{1}{g^2} \bigg[ \beta I_{AB}^{ID} - \chi_{EB}^{ID} \bigg],
\end{align}
where $I^{ID}_{AB}$ and $\chi^{ID}_{EB}$ are calculated from the covariance matrix
\begin{align}\label{gg02-id-cv}
    V_{AB}^{ID} &= \begin{pmatrix} V'\mathbb{I} & \sqrt{T'[(V')^2 - 1]}\sigma_Z \\ \sqrt{T'[(V')^2 - 1]}\sigma_Z & T'(V' + \chi' + \e_{BER}')\mathbb{I}\end{pmatrix},
\end{align}
with $\e_{BER}' = 4(\alpha')^2e_C^{ID}$.

\begin{figure}[!t]
    \centering
    \begin{tikzpicture}[
    RDET/.style={
    semicircle, draw=black!80, fill=gray!40, thick, minimum size=4mm, rotate = 270
    },
    LDET/.style={
    semicircle, draw=black!80, fill=gray!40, thick, minimum size=4mm, rotate = 90
    },
    TMSVS/.style={
    rectangle, rounded corners, draw=black!80, fill=green!15, thick, minimum size=5mm
    },
    CHANNEL/.style={
    rectangle, draw=black!80, fill=gray!5, thick, minimum size=7.5mm
    },
    QSCOVER/.style={
    rectangle, rounded corners, draw=black!80, dashed, fill=cyan!20, thick, inner xsep=12mm, inner ysep=18mm
    },
    QSDETTOP/.style={
    semicircle, draw=black!80, fill=gray!40, thick, minimum size=2.5mm
    },
    QSDETRIGHT/.style={
    semicircle, draw=black!80, fill=gray!40, thick, minimum size=2.5mm, rotate = 270
    },
    QSCHANNEL/.style={
    rectangle, draw=black!80, fill=gray!5, thick, minimum size=5mm
    },
    DISPOP/.style={
    rectangle, draw=black!80, fill=orange!20, thick, minimum size = 2.5mm
    },
    LABEL/.style={
    rectangle
    }
    ]
    \coordinate (grouping) at (4.25, -0.65);
    \node[QSCOVER](qs)[above = 0.0cm of grouping, anchor = center]{};
    
    \node[TMSVS](TMSVS){TMSVS};
    \node[LDET](aliceHet)[left = 0.5cm of TMSVS, anchor=chord center]{};
    \node[DISPOP](sqccDisplace)[right = 0.1cm of TMSVS, anchor=west]{ \scriptsize $\hat D(\Tilde{\alpha})$};
    \node[CHANNEL](channel)[right = 0.3cm of sqccDisplace]{};
    \node[QSCHANNEL](qschannelBalanced)[right = 1.25cm of channel, anchor=west]{};
    \node[QSCHANNEL](qschannelTau)[below=of qschannelBalanced, anchor=center]{};
    \node[RDET](bobHet)[right = 1.5cm of qschannelTau, anchor=chord center]{};
    \node[QSDETTOP](qsdetTop)[above = 0.5cm of qschannelBalanced, anchor=chord center]{};
    \node[QSDETRIGHT](qsdetRight)[right = 0.5cm of qschannelBalanced, anchor=chord center]{};
    \node[LABEL](channelIn)[below = 0.5cm of channel]{};
    \node[LABEL](channelOut)[above = 0.5cm of channel]{};
    \node[LABEL](qsInLeft)[left = 0.5cm of qschannelTau]{};
    \node[LABEL](qsInBelow)[below = 0.5cm of qschannelTau]{};

    \node[LABEL](AliceL)[above = 0.1cm of aliceHet.mid east]{Alice};
    \node[LABEL](BobL)[above = 0.1cm of bobHet.mid west]{Bob};
    \node[LABEL](aliceHetL)[below = 0.0cm of aliceHet.mid west]{HET};
    \node[LABEL](bobHetL)[below = 0.0cm of bobHet.mid east]{HET};
    \node[LABEL](channelCharact)[left = 0.25cm of channel.south west, anchor=north]{$\{ T, \e_0 \}$};
    \node[LABEL](modeAL)[above right = 0.0cm of aliceHet.chord center]{A};
    \node[LABEL](modeBL)[above left = 0.0cm of bobHet.chord center]{B2};
    \node[LABEL](modeB1L)[above left = 0.0cm of qsdetRight.chord center]{\scriptsize B};
    \node[LABEL](modeB2L)[below right = 0.0cm of qsdetTop.chord center]{\scriptsize B1};
    \node[LABEL](1photonL)[below = -0.1cm of qsInBelow.north]{$\ket{1}$};
    \node[LABEL](vacuumL)[left = -0.1cm of qsInLeft.east]{$\ket{0}$};
    \node[LABEL](qsTauL)[below left = -0.1cm of qschannelTau]{$\tau$};
    \node[LABEL](qsBalL)[below left = -0.1cm of qschannelBalanced]{\scriptsize 1/2};
    \node[LABEL](qsL)[below right = 0.0cm of qs.north west, anchor = north west]{QS};
    
    \draw[-] (TMSVS.east) -- (sqccDisplace.west);
    \draw[-] (sqccDisplace.east) -- (channel.center);
    \draw[->] (TMSVS.west) -- (aliceHet.chord center);
    \draw[-] (channel.center) -- (qschannelBalanced.center);
    \draw[->] (qschannelBalanced.center) -- (qsdetRight.chord center);
    \draw[-] (qschannelBalanced.center) -- (qschannelTau.center);
    \draw[->] (qschannelBalanced.center) -- (qsdetTop.chord center);
    \draw[->] (qschannelTau.center) -- (bobHet.chord center);
    \draw[-] (channelIn.north) -- (channel.center);
    \draw[->] (channel.center) -- (channelOut.south);
    \draw[-] (channel.south west) -- (channel.north east);
    \draw[-] (qschannelBalanced.south west) -- (qschannelBalanced.north east);
    \draw[-] (qschannelTau.south west) -- (qschannelTau.north east);
    \draw[-] (qsInBelow.north) -- (qschannelTau.center);
    \draw[-] (qsInLeft.east) -- (qschannelTau.center);
    
\end{tikzpicture}
    \vspace{-1.5em}
    \caption{The quantum scissor-amplified SQCC CV-QKD scheme. TMSVS denotes Alice's two-mode squeezed state, of which one mode is retained and heterodyned by Alice (HET). The other mode is displaced by $\alpha$ and sent through a lossy channel of transmissivity $T$ and excess noise $\e$, whereupon Bob passes it through a quantum scissor of gain $g = \sqrt{(1 - \tau)/\tau}$ and heterodynes the output on success.}
    \label{fig:sqcc-qs}
    \vspace{-1.5em}
\end{figure}
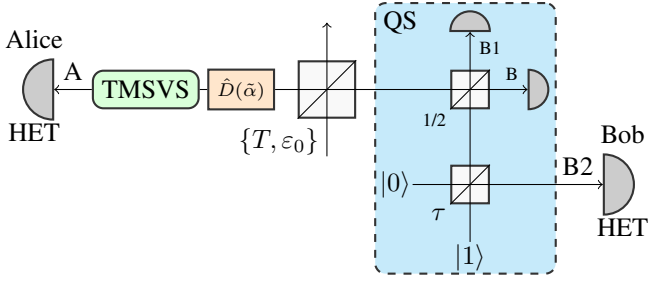

\subsection{Quantum scissor} \label{sec:sqcc-qs}

A physically-realizable approximation to the ideal NLA protocol is one utilising a first-order quantum scissor (Fig.~\ref{fig:sqcc-qs}). An $n^\text{th}$-order quantum scissor, such as those described by~\cite{winnel_generalized_2020}, perform state amplification by entangling the target state with one half of an $n$-photon resource state and performing a generalized Bell state measurement on the output modes. This effectively projects the input state onto the other half of the resource state, conditional on obtaining the correct measurement outcomes. For infinite-dimensional coherent input states, the procedure simultaneously amplifies the coherent amplitude and truncates the state to order $n$ in the Fock basis.
\begin{figure}[!t]
    \centering
    \includegraphics[width = \columnwidth]{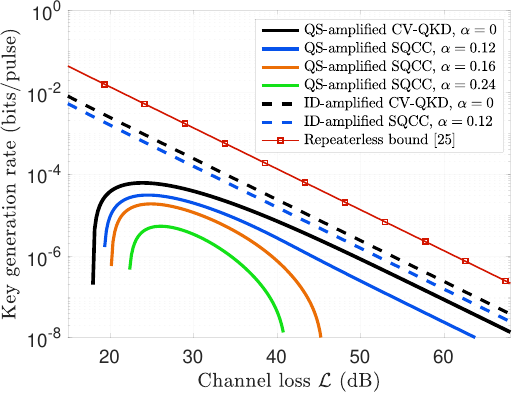}
    \caption{$\{ g, V \}$-optimised KGRs for SQCC protocols employing ideal (ID, dashed lines) and quantum scissor-type (QS, solid lines) receiver-side NLA. Key rates and channel capacities are shown as a function of channel loss $\mathcal{L}$ for thermal excess noise $\e_0 = 0.03$, phase noise $\sigma = 10^{-6}$, reconciliation efficiency $\beta = 0.95$ and BPSK orientation $\theta = 0$. In the case of $\alpha~=~0$, both protocols recover the amplified GMCS protocols described in \cite{notarnicola_long-distance_2023}. For small $\alpha$, the quantum scissor-assisted protocol scales proportionally to the repeaterless bound \cite{pirandola_fundamental_2017}, which acts as a global upper limit for quantum performance of both SQCC and non-SQCC protocols.} 
    \label{fig:pubfig_p2_x}
    \vspace{-1.5em}
\end{figure}
In the case where Bob employs a first-order $(n = 1)$ quantum scissor at the receiver, the key rate equation becomes
\begin{align}\label{KGR-GG02-QS}
    K^{QS}(g, V) = 2 P^{QS} \left[ \beta I_{AB}^{QS} - \chi_{EB}^{QS} \right],
\end{align}
for probability of success $P^{QS}$; $I_{AB}^{QS}, \chi_{EB}^{QS}$ are calculated from
\begin{align}
    V_{AB}^{QS} &= \begin{pmatrix} A^{QS}\mathbb{I} & C^{QS}\sigma_z \\ C^{QS} \sigma_z & \left[ B^{QS} + T\e^{QS}_{BER}\right] \mathbb{I} \end{pmatrix},
\end{align}
for $\e^{QS}_{BER} = 4 d^2 e_C^{QS}$. The covariance matrix $V_{AB}^{QS}$ is computed explicitly from the joint state post-scissor.

Note that while the key rate is dependent on the complex displacement parameter $\alpha_C$ as well as the gain and variance parameters $g$ and $V$, it is not especially instructive to optimise over either the displacement magnitude $\alpha$ or displacement phase $\theta$. For this reason, we restrict ourselves when optimizing to $g$ and $V$, opting to take $\theta = 0$ for simplicity.

We present the optimised key rate as a function of loss for the ideal NLA-assisted SQCC protocol in Figure \ref{fig:pubfig_p2_x}. For zero classical displacement, the SQCC protocol exactly replicates the performance of the non-SQCC protocol described in \cite{notarnicola_long-distance_2023}, scaling with loss proportionally to the repeaterless bound \cite{pirandola_fundamental_2017}. As the displacement $\alpha$ increases, the associated key rates exhibit the same ideal scaling behaviour, though at a reduced magnitude. This linear scaling with loss for $T \ll 1$ occurs when the optimal gain is chosen such that $g^2 \propto 1/T$ and so the quantity $g^2 T$ becomes constant; under these conditions, the protocol assumes an effective transmissivity $T_\text{eff}$ which saturates at high loss. This has the effect of saturating $I_{AB}^{Id}$ and $\chi_{AB}^{Id}$, and so $K \propto 1/g^2 \propto T$. For the case where $\alpha = 0.12$, the optimal gain becomes $g \sim 731$ at 60 dB, with $g^2 T$ and $T_\text{eff}$ saturating to $0.53$ and $0.31$ respectively. As a consequence of the saturated effective transmissivity, the classical performance of the ideal NLA-assisted protocol asymptotically approaches a constant value, with the bit-error rate saturating to $e^{Id}_C = 0.475$ beyond $20$~dB for e.g. $\alpha = 0.12$. Thus, the ideally-amplified protocol significantly improves upon the non-amplified protocol in quantum performance, demonstrating no upper bound on the loss tolerable by the protocol, and provides nonzero classical throughput for all $\mathcal{L}$.
\begin{figure}[!t]
    \centering
    \begin{tikzpicture}[
    RDET/.style={
    semicircle, draw=black!80, fill=gray!40, thick, minimum size=4mm, rotate = 270
    },
    LDET/.style={
    semicircle, draw=black!80, fill=gray!40, thick, minimum size=4mm, rotate = 90
    },
    TMSVS/.style={
    rectangle, rounded corners, draw=black!80, fill=green!15, thick, minimum size=5mm
    },
    CHANNEL/.style={
    rectangle, draw=black!80, fill=gray!5, thick, minimum size=7.5mm
    },
    QSCOVER/.style={
    rectangle, rounded corners, draw=black!80, dashed, fill=cyan!20, thick, inner xsep=3mm, inner ysep=3mm
    },
    BOBCOVER/.style={
    rectangle, rounded corners, draw=black!80, dashed, fill=red!10, thick, inner xsep=20mm, inner ysep=19mm
    },
    TDET/.style={
    semicircle, draw=black!80, fill=gray!40, thick, minimum size=4mm
    },
    QSCHANNEL/.style={
    rectangle, draw=black!80, fill=gray!5, thick, minimum size=5mm
    },
    DISPOP/.style={
    rectangle, draw=black!80, fill=orange!20, thick, minimum size = 2.5mm
    },
    LABEL/.style={
    rectangle
    }
    ]
    \node[TMSVS](TMSVS){TMSVS};
    \node[LDET](aliceHet)[left = 0.3cm of TMSVS, anchor=chord center]{};
    \node[DISPOP](sqccDisplace)[right = 0.1cm of TMSVS, anchor=west]{ \scriptsize $\hat D(\pm \tilde \alpha)$};
    \node[CHANNEL](channel)[right = 0.15cm of sqccDisplace]{};
    \node[BOBCOVER](bobcover)[right = 0.1cm of channel, anchor = west]{};
    \node[QSCHANNEL](classicalBS)[right = 0.75cm of channel, anchor=west]{};
    \node[DISPOP](reverseDisplace)[right = 1.0cm of classicalBS.east, anchor=center]{ \scriptsize $\hat D(\mp\sqrt{tT} \tilde \alpha)$};
    \node[QSCOVER](qscissor)[below = 0.3cm of reverseDisplace]{QS};
    \node[RDET](bobHet)[right = 0.6cm of qscissor, anchor=chord center]{};
    \node[TDET](classicalHet)[above = 0.5cm of classicalBS, anchor=chord center]{};
    \node[LABEL](channelIn)[below = 0.5cm of channel]{};
    \node[LABEL](channelOut)[above = 0.5cm of channel]{};
    \node[LABEL](classicalBSIn)[below = 0.5cm of classicalBS]{};

    \node[LABEL](AliceL)[above = 0.1cm of aliceHet.mid east]{Alice};
    \node[LABEL](BobL)[above = 2.1cm of bobHet.mid west]{Bob};
    \node[LABEL](aliceHetL)[below = 0.0cm of aliceHet.mid west]{HET};
    \node[LABEL](bobHetL)[below = 0.0cm of bobHet.mid east]{HET};
    \node[LABEL](channelCharact)[left = 0.25cm of channel.south west, anchor=north]{$\{ T, \e_0 \}$};
    \node[LABEL](qsBalL)[below left = -0.1cm of classicalBS]{t};
    \node[LABEL](classicalHetL1)[right = 0.0cm of classicalHet]{HET};
    \node[LABEL](classicalHetL2)[above = 0.0cm of classicalHet]{\hspace{0.25cm}\scriptsize$\{\pm\sqrt{(1-t)T} \tilde \alpha \}$};
    
    \draw[-] (TMSVS.east) -- (sqccDisplace.west);
    \draw[-] (sqccDisplace.east) -- (channel.center);
    \draw[->] (TMSVS.west) -- (aliceHet.chord center);
    \draw[-] (channel.center) -- (classicalBS.center);
    \draw[->] (classicalBS.center) -- (classicalHet.chord center);
    \draw[-] (classicalBS.center) -- (reverseDisplace.west);
    \draw[-] (reverseDisplace.south) -- (qscissor.north);
    \draw[->] (qscissor.east) -- (bobHet.chord center);
    \draw[-] (channelIn.north) -- (channel.center);
    \draw[->] (channel.center) -- (channelOut.south);
    \draw[-] (channel.south west) -- (channel.north east);
    \draw[-] (classicalBS.south west) -- (classicalBS.north east);

\end{tikzpicture}
    \vspace{-0.5cm}
    \caption{The amplified SQCC scheme with dual measurement. Bob uses a beamsplitter of transmissivity $t$ to extract a small portion of the signal, from which he measures the classical bit of the signal. Given the classical measurement, the remaining signal is re-displaced to zero mean before amplification and measurement.}
    \label{fig:sqcc-qs-bs}
    \vspace{-1.5em}
\end{figure}
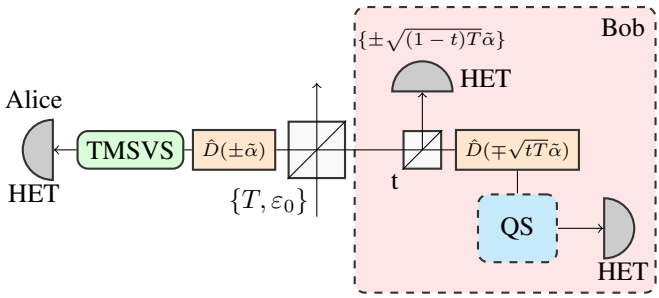

Figure \ref{fig:pubfig_p2_x} also presents optimised key rates as a function of loss for the SQCC protocol amplified by the physically-realizable quantum scissor NLA. The protocol again replicates~\cite{notarnicola_long-distance_2023} at $\alpha = 0$, with the KGR curve demonstrating a nonzero keyrate at any channel loss for zero classical throughput in the high-loss limit. Importantly, we also obtain optimal scaling for nonzero $\alpha$ in the low-$\alpha$ regime ($\alpha~\leq~10^{-1}$) concurrently with an asymptotically saturating bit-error rate, indicating that in this regime our protocol is capable of delivering both secret-key and classical information for any channel loss. This occurs in an identical manner to the ideally-amplified protocol, where the amplifier generates an effective channel transmissivity which saturates for high loss; for $\alpha = 0.12$ the optimal gain becomes $g \sim 22$ at 60~dB, with $g^2 T$ and consequently $T_\text{eff}$ saturating to $5 \times 10^{-4}$. Thus, the physically-amplified protocol exhibits the same scaling with loss for $T \ll 1$ as in the ideal NLA in both quantum and classical performance, reinforcing the findings of \cite{notarnicola_long-distance_2023}, though we note that the protocol can only achieve this optimal scaling for $\alpha \leq 10^{-1}$. Above this limit, the protocol can only deliver a nonzero secure key rate over a finite range of losses. Further, unlike previous protocols, the quantum-scissor assisted protocol cannot generate a secure key in the large-$\alpha$ regime ($\alpha \geq 0.3$) as a result of the truncation performed by the quantum scissor and the excess noise contribution from the classical signal.
For comparison, we include in Fig.~\ref{fig:pubfig_p2_x} the repeaterless bound \cite{pirandola_fundamental_2017}, which represents the ultimate rate for two-way quantum communications over a lossy bosonic channel. However,
broader insights into the information-theoretical behaviour of the presented SQCC protocol could be gained by implementing a trade-off coding framework. This framework is discussed in more detail later (see the Discussion section).
 
Additionally, the quantum scissor-assisted protocol, while being the first example of a physically-realizable SQCC protocol with simultaneous nontrivial classical throughput and nonzero key generation rate at arbitrary loss, is constrained by the physical limitations of the scissor device. Specifically, the truncation of the output state to first order in the Fock basis bounds above the coherent-state amplitude that Bob can process without loss of information. This restriction increases the bit-error rate of the signal relative to the ideal-amplifier protocol, to approximately $e_C^{QS} = 0.496$ for $\alpha = 0.24$ and $e_C^{QS} = 0.498$ for $\alpha = 0.12$. Proper use of contemporary classical error-correction codes, however, would be sufficient to make classical communications over such a channel viable, while also improving the quantum portion of the combined signal by reducing the postprocessing noise $\e_{BER}$.

\begin{figure}[!t]
    \vspace{-2em}
    \centering
    \includegraphics[width = \columnwidth]{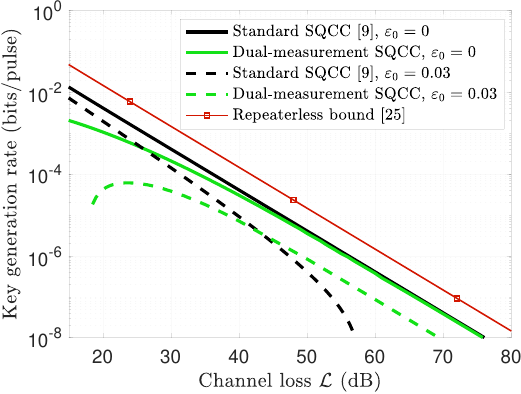}
    \caption{Key generation rate as a function of channel loss $\mathcal{L}$ for the novel dual-measurement SQCC protocol proposed in Section \ref{sec:sqcc-bsqs}. The protocol performance is compared with the standard non-amplified SQCC protocol \cite{qi_simultaneous_2016} in the ideal case of $\alpha = 10^{12}, \ \beta = 0.95$ and $\sigma = 0$ for $\e_0 = 0, \ 0.03$. Each protocol is optimised over all free parameters, namely $\{V, g, t\}$ and $\{V\}$ respectively. Classical bit-error rates are arbitrarily close to zero in all cases.}     
    \label{fig:sqcc_bsqs}
    \vspace{-1.0em}
\end{figure}

\subsection{Quantum scissor and beamsplitter} \label{sec:sqcc-bsqs}
To circumvent the low classical throughput associated with the quantum scissor-assisted protocol, we propose a new scheme, shown in Figure \ref{fig:sqcc-qs-bs}. In this scheme, the received signal is split into two modes via a beamsplitter of transmissivity $t \sim 1$. By performing a heterodyne detection on the weaker mode, Bob extracts the displacement of the SQCC signal and subsequently the encoded classical bit, which he can then use to reverse the classical encoding on the remaining mode. This reduces the SQCC protocol to zero-mean Gaussian-modulated coherent-state CV-QKD, which Bob amplifies by a quantum scissor prior to heterodyne measurement. 

The motivation of this scheme is that in separating the quantum and classical measurements, Bob can simultaneously gain superior classical throughput via a larger classical displacement $\alpha$, while also exploiting noiseless amplification for superior quantum throughput, which works optimally for $\alpha \simeq 0$. In the high-displacement, low-noise limit ($\alpha \longrightarrow \infty$, $\sigma \longrightarrow 0$) we expect this novel protocol to achieve perfect classical throughput, since for arbitrarily large $\alpha$ the fraction extracted by the beamsplitter retains a sufficiently high signal-to-noise ratio to reduce the bit-error rate arbitrarily close to zero. Given perfect classical measurement, Bob can reverse Alice's large classical encoding on the signal to obtain a zero-mean CV-QKD signal without introducing excess noise. From earlier results (Fig.~\ref{fig:pubfig_p2_x}), we expect the application of noiseless amplification to this signal to produce a protocol which scales proportionally to the repeaterless bound, achieving near-optimal quantum throughput. This behaviour is demonstrated in Figure \ref{fig:sqcc_bsqs}. While the low-noise limit is not yet fully realistic, the requirement of a large classical displacement makes this protocol ideal for integration with classical FSO communications networks, where beam energies are much higher (${\sim}10$~mW) than in QKD networks.

\section{Discussion}
We close our work with a short discussion on the optimal bounds related to our calculations. We have investigated here the performance of SQCC using the same signal and explored the tradeoff in energy dedicated to each communication component.
A major motivation for such combined communication is the scenario wherein a single physical transmitter on board a satellite is utilised for both types of communication. This scenario leads to simpler and more pragmatic satellite designs in that, for example, independent radio transmitters would no longer be required for classical communications. In this context, we have explored the performance of classical communications combined with QKD over a  noisy and lossy bosonic channel, together with the implicit assumption that all classical communications required for QKD would be encapsulated in the classical portion of our protocol (assuming two-way simultaneous communications between sender-receiver pairs). Our calculations illustrate the tradeoff in secret-key generation rates as we allow for more classical communications.
Considering QKD alone, bounds on rates (per channel use) for the lossy bosonic channel are known \cite{takeoka_fundamental_2014, garcia-patron_reverse_2009} and are defined by $g[(1+T)N_m/2]-g[(1-T)N_m/2]$ for Shannon entropy $g$ of a zero-mean variable with geometric distribution and mode photon number $N_m$ \cite{garcia-patron_reverse_2009}; in the limit of large $N_m$ this bound approaches $\log_2(1+T)/(1-T)$. Improved analysis leads to the tighter bound of $-\log_2(1-T)$ \cite{pirandola_fundamental_2017}. However, these bounds implicitly assume unbounded two-way classical communication, and therefore are not directly related to our scenario.
We have not explored the use of coding within our calculations, but rather explicitly followed the bit error rate of the classical communication and the QKD rates of the quantum communications. A more sophisticated approach would involve both the inclusion of classical codes and quantum error correction codes for the classical and quantum portions of the protocol, respectively. Previous works have investigated formal capacity bounds within this context \cite{wilde_information_2012}, and these bounds are more relevant to our scenario, with combined classical-quantum tradeoff codes found to be dependent on the both number of photons utilized  and the fraction of photons dedicated to the quantum or classical signal. We anticipate future work in the area of SQCC will explore pragmatic coding designs that meets the formal capacity bounds of \cite{wilde_information_2012}. However, we note that in the context of the calculations reported here, we would not be interested in the high-energy limit of these bounds but rather in determining the optimal classical-quantum coding scheme that meets a given energy constraint on the carrier state containing both classical and quantum information. The results reported here indicate the tradeoff in performance between classical and quantum communications that would be anticipated in such an energy-constrained circumstance (whilst optimizing some parameters to meet quality of service requirements). We also note that the tradeoffs shown are available immediately - they do not require quantum computing devices at the receiving nodes to implement quantum error correction.

\section{Conclusion} \label{sec:conclusion}
In this work, we assessed the viability of combined quantum-classical protocols in the context of realistic satellite-based communications networks, which are constrained to operate in the low-transmission energy regime. We presented results detailing the minimum pulse intensity requirements of an SQCC signal to achieve a given pair of quantum and classical quality-of-service thresholds, given a free space optical channel of fixed transmissivity and thermal noise.

We then discussed novel SQCC protocols utilising both idealized and physically-realizable noiseless amplification at the receiver end and derived optimised protocol performances for realistic systems. Our results demonstrate that receiver-side noiseless amplification permits SQCC protocols to deliver nonzero secret-key rates and nonzero classical channel capacities at any large loss, something that cannot be replicated by equivalent non-amplified protocols in the small-classical-displacement regime. Furthermore, we presented a novel scheme exploiting dual measurement of SQCC signals in tandem with noiseless amplification and demonstrated that such a scheme achieves in-principle near-perfect classical and quantum throughput. Our results illustrate a feasible pathway to integration of quantum communication with classical FSO systems.

\section{Acknowledgement}
The Australian Government supported this research through the Australian Research Council’s Linkage Projects funding scheme (Project No. LP200100601). The views expressed herein are those of the authors and are not necessarily those of the Australian Government or the Australian Research Council. Approved for Public Release: NG24-0370.

\begin{appendices}
\section{Derivation of ideal NLA results} \label{sec:appIDNLA}
For a GMCS protocol with operational parameters $\{ V, T ,\e \}$, introducing an ideal NLA operation $g^{\hat n}$ at the receiver side prior to Bob's measurement is equivalent to a non-amplified GMCS protocol with transformed operational parameters \cite{blandino_improving_2012}
\begin{align}
    \{V', T', \e' \} \Longrightarrow 
    \begin{cases}
        V' = V^{Id}(\lambda, T, \e, g) \\
        T' = T^{Id}(V, T, \e, g) \\
        \e' = \e^{Id}(V, T, \e, g).
    \end{cases}
\end{align}
We utilise the method described in \cite{blandino_improving_2012} to determine the equivalent non-amplified channel in the SQCC case. However, we must carefully consider how the effective channel transforms, both in light of the additional excess noise terms and the new free parameter $\alpha$, describing the magnitude of the classical displacement, which governs the classical performance. We find a set of non-amplified parameters $\{ \alpha', V', T', \e_0' \}$ that provide equivalent performance to an amplified SQCC protocol of parameters $\{ \alpha, V, T, \e_0 \}$ and gain $g$:
\begin{align} \label{idnla-sqcc-params}
    \{\alpha', V', T', \e_0' \} \Longrightarrow 
    \begin{cases}
        \alpha' = \alpha^{ID}_{SQCC}(\alpha, V, T, \e, g) \\
        V' = V^{ID}_{SQCC}(\alpha, V, T, \e, g) \\
        T' = T^{ID}_{SQCC}(\alpha, V, T, \e, g) \\
        \e_0' = \e^{ID}_{SQCC}(\alpha, V, T, \e, g).
    \end{cases}
\end{align}
We observe that for the non-SQCC case, where $\alpha = 0$, we recover $\{ \alpha', V', T', \e_0' \}~=~\{ 0, V^{ID}, T^{ID}, \e^{ID} \}$. Thus the equivalent channel parameters $\{ V', T', \e' \}$ for $\alpha = 0$ describe the same channel as a non-SQCC GMCS protocol boosted by an ideal NLA, as expected. Similarly, we see that for the non-amplified protocol with $g = 1$ we obtain $\{ \alpha', V', T', \e_0' \}~=~\{ \alpha, V, T, \e_0 \}$, i.e. the protocol without amplification returns the original scheme. 

Given \eqref{idnla-sqcc-params}, the BER of Bob's signal is computed by
\begin{align}
    e_C^{ID} &= \frac{1}{2} \text{erfc} \left( \sqrt{\frac{T'(\alpha')^2}{2 B^{ID}}} \right),
\end{align}
for $B^{ID} = T'[V' + \chi']$ and $\chi' \equiv \frac{1 - T'}{T'} + \e_0' +(\alpha')^2\sigma$.

\section{Derivation of quantum scissor results} \label{sec:appQS}
Bob implements a first-order quantum scissor at the receiver side in the following way. He prepares two auxiliary modes, containing a single photon and vacuum respectively, which he coherently combines on a beamsplitter of transmissivity $\tau$ to generate the entangled state $\ket{\varphi} = \sqrt{1-\tau}\ket{01} - \sqrt{\tau}\ket{10}$. He then mixes one half of this entangled state with the Gaussian state he received from Alice, which is described by the covariance matrix \eqref{tmsvs-cv}, on a balanced beamsplitter. Lastly, he executes the following POVMs on the two output modes:
\begin{align}
    \left\{ \hat \Pi_1 = \sum_{n = 1}^\infty \ket{n}\bra{n}, \ \hat \Pi_0 = \mathbb{I} - \hat \Pi_1 \right\}.
\end{align}
These measurement elements represent on-off detection of a quantum detector of unity efficiency. Successful operation of the scissors, where the amplified input state is truncated and teleported onto the remaining half of the resource state, is heralded by Bob measuring precisely one `on' ($\hat \Pi_1$) measurement and precisely one `off' ($\hat \Pi_0$) measurement on the two detectors. When this occurs, the state is amplified by the gain coefficient $g = \sqrt{1 - \tau/\tau}$ \cite{ralph_nondeterministic_2009} with an overall probability
\begin{equation}
    P^{QS} = \frac{4 e^{-\frac{\alpha ^2}{2 S}} \left[\tau  \left(\alpha ^2 -2 S\right)+S^2\right]}{S^3} - \frac{2 (1-\tau) e^{-\frac{\alpha ^2}{2 R}}}{R},
\end{equation}
for $R = 2 + T(V + \e - 1)$ and $S = 2 + R$. Bob then performs a heterodyne measurement on the output port of the scissor. 

We note that since the output of the scissor is necessarily non-Gaussian, the security of the key rate given in Section II is no longer guaranteed \cite{garcia-patron_unconditional_2006}. To circumvent this issue, we first observe that extremity properties of Gaussian states \cite{leverrier_continuous-variable_2011} ensures that Eve's accessible information on Bob's non-Gaussian state is upper-bounded by the Holevo information obtained for a Gaussian state with identical first and second moments. Secondly, we observe that the mutual information $I_{AB}$ is lower-bounded by the mutual information for the equivalent Gaussian state \cite{wolf_extremality_2006}. We therefore compute a lower bound on the key rate using the equivalent Gaussian state with identical mean and covariance to Bob's state post-amplification.

Calculation of the key rate equation for the case in which the SQCC protocol is amplified by a quantum scissor at the receiver end follows identically to the quantum scissor-assisted non-SQCC protocol described by \cite{notarnicola_long-distance_2023}, with the exception that the success probability and covariance matrix are now dependent on the complex displacement $\tilde \alpha$ used to encode the classical information via a BPSK modulation ($\tilde\alpha~\in~\{\pm\alpha e^{i\theta}\}$). We determine the bit-error rate of Bob's classical signal via
\begin{align} \label{BER-SQCC-QS}
    e_C^{QS} &= \frac{1}{2} \text{erfc} \left( \sqrt{\frac{d_{QS}^2}{2 B^{QS}}} \right),
\end{align}
where $d_{QS}$ and $B^{QS}$ are the mean displacement and variance of Bob's received mode calculated explicitly from the quantum state output by the scissor:
\begin{align}
    d_{QS}^2 &= \frac{16 \alpha ^2 (1-\tau) \tau  R^2 S^2 e^{\frac{\alpha ^2}{R}}}{\left(2 R \left[ \tau  \left(\alpha ^2-2 S\right)+S^2\right]e^{\frac{\alpha ^2}{2 R}}-S^3 (1-\tau) e^{\frac{\alpha ^2 }{2 S}}\right)^2}\\
    B^{QS} &= \frac{\left( \splitdfrac{- 32 \alpha ^2 R (1 - \tau) \tau e^{\frac{\alpha ^2 }{2 R}} \cos ^2\theta - 3S^4 (1 - \tau) e^{\frac{\alpha ^2 }{S}}}{ + 2 R S e^{\frac{\alpha ^2 (R+S)}{2 R S}} \left[\tau  \left(\alpha ^2 -2 S\right) - S^2 (2 \tau - 3 )\right]} \right)}{2 R S e^{\frac{\alpha ^2 (R+S)}{2 R S}} \left[\tau  \left(\alpha ^2 -2 S\right)+S^2\right]-S^4 (1 - \tau) e^{\frac{\alpha ^2 }{S}}}\label{eq:qs_bvariance}
\end{align}
As the truncated state may be highly non-Gaussian, it is not necessarily appropriate to use the above Gaussian formulation to calculate the BER. However, we estimate that in the low-amplitude, low-squeezing regime the state Bob receives is sufficiently Gaussian so as to make Eq. \eqref{BER-SQCC-QS} a suitable approximation to the true BER. The remaining covariances of Alice and Bob's joint quantum state, $A^{QS}$ and $C^{QS}$, are calculated similarly but have been excluded for brevity.
\end{appendices}

\bibliographystyle{IEEEtran}
\bibliography{references}{}
\clearpage

\end{document}